# A Comparative Study on Self-Organization in Wireless Sensor Networks

Michael Simon
School of Electrical Engineering
Ontario Institute of Technology
Email: michael.simon@ontario.com

Salwa M. Din
York University
Toronto , Canada
Email: salkammd@my.yorku.ca

Raja Jamal Chib
WBAN-Sensor Academy
San Francisco, California 96078–3391
Email: rajaJamal@wban.net

*Abstract*— With advancements in microelectromechanical systems, low-power integrated circuits, and wireless communications, wireless sensor networks (WSNs) have become increasingly significant [1][2]. These distributed networks enable efficient resource utilization and open doors to numerous applications, including personal healthcare, home automation, environmental monitoring, industrial automation, and defense surveillance. However, WSNs are susceptible to environmental factors in their deployment areas and may suffer damage. In such cases, the network must be reconfigured or repaired. To address these challenges and adapt to resource constraints, WSN mechanisms must exhibit self-organizing capabilities. For instance, in tasks like allocation, cooperative communication, and dynamic data collection, self-organization enhances the efficiency and robustness of WSNs across the application, network, and physical layers.

## I. INTRODUCTION

Self-organization features in a wireless sensor network manage or optimize the network in a specified environment. The application layer of wireless sensor networks can better perform collaborative signal processing, surveillance, and provide results to raise awareness and self-maintenance using bio-inspired self-organization. Although self-organizing applications help WSNs obtain information proactively and to some degree serve as preventive security, monitoring and maintaining the surveillance of WSNs are WSN applications themselves. An effective performance evaluation is required to be developed for every application of a wireless sensor network. This performance evaluation measures and improves its network operations; sensor networks have a low-power ad hoc network on a small scale having hundreds of sensor nodes and a limited energy resource.

## II. FUNDAMENTALS OF WIRELESS SENSOR NETWORKS

Wireless Sensor Networks (WSNs) comprise collections of small, low-power physical devices capable of sensing the environment, acting on it by controlling or actuating things in physical spaces, and communicating their observations and/or the results of their computations to other devices. The primary components of WSNs include sensors, communication systems, and actuators. WSNs have a wide spectrum of applications, including health care, industrial automation, home automation, traffic monitoring, habitat monitoring, and other applications like transportation and logistics, precision agriculture, remote sensing, earthquake and seismic monitoring, military, security, and surveillance applications, and the monitoring of industrial applications[3].

WSNs are capable of sensing, processing information, and taking action based upon the gathered information. They use distribution protocols; therefore, they can easily be deployed and scalable systems for users. Each node uses its unique identifier for packet routing [4]. WSNs can use selfconfiguring access channels, operate on low processing power, and generally operate on the characteristics of wireless communication. Self-organizing is important for wireless sensor networks because it provides features that are very difficult to achieve with static configuration of the sensor nodes. It makes the network continuous and collaborative, as shown, and allows the network to maintain routing capability and the ability to detect and correct faults. It is also an effective method to extend network scalability. Network topology can change dynamically to adapt to environmental or application demands or in the case of individual sensors or methods of communication falling victim to failures for a range of possible reasons [5].

### A. Definition and Components

A wireless sensor network (WSN) is a network comprised of a large number of small sensor nodes transmitting data to one or more data acquisition systems via a wireless communication architecture. WSNs can be used for a variety of applications, including infrastructure monitoring, environmental studies, defense applications, and healthcare. Data generated by sensor nodes can be collected, aggregated, and disseminated in a multi-hop fashion; a data-intensive application can generate a plethora of data. Sensor nodes have limited resources, such as computing power, memory, and energy, and are joined by the nature of their environment, which is transient or typically inaccessible. Data acquisition systems can use a cellular or other connection to transmit the data to the application's end user [6].

The primary components of WSNs are sensor nodes, a communication protocol that controls radio frequency (RF)

communication, electronic hardware and software that manage sensors, and data and other software applications running on computers in the network or off-site computers. The nodes in WSNs have responsibilities that include data acquisition, data aggregation, power management, and radio communication. Many other services could be provided by nodes, such as preprocessing data for acceptable propagation rates, eliminating extraneous or redundant data, performing simple mathematical calculations, and using data to make decisions based on protocols. In order to be aware of the next links that need to be taken into account in the data transfer chain, sensor nodes must be "aware" of each other's potency. These can include: - The medium that the network uses to communicate; - Hardware that uses electronics and software; - Protocols, which instruct others on how to pass data from user to user accurately, securely, and without loss of data. These essential components must be understood before assessing the performance of a WSN [7].

*B. Applications*

Wireless sensor networks (WSNs) have several applications, including environmental monitoring, healthcare systems, agricultural systems, industrial automation, pollution detection, object tracking, waste management systems, stations for collecting meteorological and environmental data, and smart cities. In smart cities, sensor networks are utilized to ensure the safety and security of citizens, control traffic efficiently, and gather information related to different systems and services. Smart cities ensure both technological and economic advancements in the urban realm. WSNs have been utilized in automotive systems, particularly to anticipate foreseeable problems such as fuel consumption and air pressure. They have also been used in the automotive industry for simple functionalities such as controlling the airbag system. The demand for real-time monitoring and automatic data collection has promoted the use of WSNs. One more example is an early warning system that utilizes WSNs and a variety of sensors to detect the measures of an embankment and assess their applicability in detecting slip surfaces in an embankment. The usage of WSNs in these applications ranges over a variety of topics, serving different purposes such as control and monitoring for dams, oil and gas, and energy applications. In agriculture, WSNs are mainly used for controlling the growth of unexpected plants away from the main area, such as the growth of weeds. WSNs are used to collect fruit quality evaluations in a short time with a wireless sensor camera. The systems rely on the geographical sensors and communication devices' long-lived nature. Currently, more studies are focused on monitoring, smart work, and providing valuable information about positioning equipment and environmental content evaluations. Recent use is for the technology of data analysis. In environmental concerns, wireless sensors are used to measure water quality in streams and lakes as an early warning network if any poisonous effects are detected. Moreover, in some cases, they may also detect the pollution present in the water. Hence, four applications have been addressed [8]. Concerning these applications, self-configuration, self-healing, self-optimization, and self-protection are discussed. In the Internet of Things section, the use of WSNs to construct smart applications is presented, which integrate sensor networks into daily life and impact everyone's life. The applications discussed are smart cities, road traffic in smart cities, smart grids, and environmental monitoring in smart cities. The recent trends and emerging areas of WSN research and applications with wireless communication technologies are burgeoning.

*C. Self-Organizing Features in Wireless Sensor Networks*

Self-organization, similar to the concept of swarm intelligence, is a bottom-up mechanism useful in wireless sensor networks for adapting the networks to the changing environment, policies, and capabilities. In this decentralized manner, which is sometimes attributed as an emergent property, WSN nodes can efficiently collaborate with other nodes as fast as necessary to meet the objectives, regardless of the number of entities involved [9]. Thus, this type of performance also ensures that the sensor network does not reduce its performance when the demands and dynamics of the events to be sensed are growing. Due to these advantages of selforganizing features, for different demands of WSN nodes and environments, many types of these features have been proposed, and their performance can be analyzed with the specific behavioral features of WSNs [10].

In dynamic and resource-constrained environments, such as in logistics and natural disaster supervision, a large scale of sensors can be deployed to monitor and sense data for more efficiency and greater coverage. Thus, operational systems in such dramatically changeable dynamics, urgent requirements, and diverse brands and types of sensors need to provide quick decisions and not reduce their performance when facing more flooding concerns. Additionally, the networks should be scalable to be able to abruptly increase the number of deployed sensors with a considerably low amount of configuration. Therefore, self-organizing principles can improve WSNs in these respects. In WSNs, self-organizing features are invented to allow nodes to self-configure based on the sensor and events' characteristics, networks' topology and application, and produce favorable arrangements of resources to achieve high performance and meet the predefined objectives of the system. The main goals of developing WSNs with selforganizing features are to amplify the network adaptability, efficiency, and scalability to the external environment changes by utilizing sensor context information while maintaining low maintenance costs. By having grown self-configuration capability with local minimalism and autonomic features, WSN networks globally drive the network's self-organization to the minimum overhead.

*1) Definition and Importance:* There is not yet a clear definition of the term "self-organizing feature" within the community. We adopt the term from macroscopic physical systems like social insects, which collectively give rise to macroscopic properties without any centralized control. Similarly, a selforganizing feature is any complex functionality that results from adaptation between the system constituents only. The main constituents of wireless sensor networks are, of course, the sensors. Especially useful are self-organizing features that lead to increased network adaptability, such as fault tolerance, resilience, and reliability, which are useful when a small, slow maintenance crew is looking after large and sporadically unreachable areas. An example of such can be adaptively selfoptimized multi-hop data transport that provides resilience in the possibly "static-free" network due to node mobility [11] . That is, the feature helps in automatically exploiting transience in the network topology, which possibly connects otherwise isolated segments.

Self-organizing is also a key enabler in delivering the desirable scalability of wireless sensor networks and mitigating the energy constraints by providing localized communications: each node only communicates with a few of its direct neighbor nodes, whose identity may be discovered in a fully decentralized way using a self-organizing algorithm; different from centralized approaches, all necessary global information is learned solely from local interactions, and there is no need for a dedicated hierarchy and dedicated energy-intensive nodes that control every conversation prior to their occurrence. Having localized communication via self-organization in large networks allows several hundreds or thousands of sensor nodes to efficiently communicate fairly reliably in a decentralized manner without creating long-living bottlenecks throughout the network [12]

*2) Types of Self-Organizing Features:* To enhance system performance, various self-organizing features are utilized in sensor networks. Categorized based on their objectives, these features include clustering techniques that facilitate balancing node energy and providing an efficient transmission approach; dynamic topology adjustment to increase robustness; and routing protocols to diversify communications that work around affected paths. Features that involve all or almost all of the aforementioned types include handoff and coveragereinforcement algorithms. The handoff mechanism encounters the dead-end problem, which pertains to the unavailability of any other neighbor for a node or a sub-cluster to move to, and keeps a node or a sub-cluster in a location with high interference, while the coverage-reinforcement area lies close to it. These techniques aim to reduce the effect of a moderate or minor fault on the overall quality of the network [13] . The ultimate goal of self-organizing properties is to improve overall network performance by the collaboration of prospective stakeholders toward any changes to existing circumstances as well as the utilization of existing features. The ability to cooperate is contingent upon individual goals or targets, and when individual goals and targets concur, overall network efficiency is enhanced. Depending on their primary concern, these methods can be regarded as cluster formation, cluster maintenance, and multi-hop wireless communication energy-saving mechanisms. Each type indirectly affects the other two. In PEGASIS, clustering is partially associated with cluster formation and cluster maintenance; therefore, it can also be labeled as a multi-hop environment, given that the aggregate values all pass through the chain of nodes. It is evident that these algorithms have an indirect association among their attributes [14] .

### III. Performance Metrics for Evaluation

We are interested in evaluating the performance of wireless sensor networks according to the most important performance metrics that can also be used as criteria for analysis. Our main attention in this respect is on the evaluation of self-organizing features, which are to be put at the heart of such networks. Performance metrics such as reliability, scalability, energy consumption, or time delay are the first choice when metrics for a system's performance have to be specified. Essentially, these metrics represent the foundation for the development of simulative case studies or real practical implementations. Networks are often compared through their metrics. New features of a system are usually assessed with respect to the current ones.

Common performance metrics with respect to the evaluation of a wireless sensor network are analyzed below. Energy consumption is a fundamental issue for many wireless sensor network applications. Thus, we are interested in networks consuming extremely small quantities of energy; that is, networks based on the self-organizing features discussed being ultralow power. In a complex system, the loss of a great number of sensor nodes, which can lead to an incomplete coverage of a region under surveillance or to an undesirable increase in maintenance operations, can be critical. Reliability is the capability of the system to exhibit behavior that is completely predictable, in an autonomous and consistent manner for a stipulated period of time. A more robust representation of the coverage quality of the nodes, however, could be given by metrics such as the number of connected sensor nodes remaining after random node failures occur; here, connectivity could be interpreted as the probability for two sensor nodes to be able to establish communication [15].

*1) Reliability:* Reliability is a fundamental performance metric of wireless sensor networks and indicates the probability that a system performs its task without errors. The three most important aspects of a reliable network are data integrity, data availability, and avoiding data loss. These issues have so far been addressed by ensuring the reliability of the network in which embedded node processors are organized. To avoid data loss, redundant pathways are usually provided by ensuring that a number of sensor nodes can communicate with a number of different field-gateway nodes or a number of completely blown-up routers. In a data acquisition application with huge amounts of data to be processed, user satisfaction generally does not really depend on real-time data. Hence, in this field, high-reliability systems have to be self-organizing. The user functions and object-relational mapping functions are to adapt to the sensor network and the non-reliable selforganized middle network between the sensor and the data processing network, which cannot allow the user functions to be self-organizing [16].

In a sensor network, the intake and transport of sensory data to their original users is known as the real-time transport demand, because sensory data will lose its value after some time. However, the system has to provide a very high level of reliability to be able to guarantee that the results are always available for the user precisely when needed. In a world where failures of sensor nodes in the network can happen all the time due to reasons such as temperature, humidity, and vibration that the node is designed for, anti-reliability is called reliability. The degree of reliability, as well as availability, degrades due to both environmental conditions, gradually reducing the number of working sensor nodes in the network, and the domain integrated condensation with momentary node dropout due to constraints during a shorter period in the propagation of, among other things, the wake-up signals. This in particular leaves the demand of the network model to maintain redundancy if the network is not dominated by sink traffic.

*2) Scalability:* Scalability is one of the most critical performance metrics, which determines the capacity for growth and indicates the limits of the wireless sensor network. Scalability indicates the sensor network's ability to provide services to an increasing number of users. In wireless sensor networks, scalability refers to the network's ability to increase its capacity as the number of nodes increases; the geographic area covered by the sensor network expands, the maximum number of links in the network, and the required diameter of the network in order to secure point-to-point communication and fault tolerance. One of the prospective methods used to preserve network scalability is to exploit self-organizing capabilities. Self-organizing wireless sensor networks can consist of a large number of inexpensive, low-power devices linked directly or through other sensor nodes to one or more sink nodes. These organizations start as original networks with sensors added or subtracted as the environment dictates [17]. The channel reservation processes in coordination with multi-channel operation and dynamic resource allocation are mainly related to scalability issues.

Scalable and efficient self-organizing mechanisms in wireless sensor networks should then guarantee robust operation with an increasing number of nodes experiencing a decreasing service quality. The quality of service can be thought of as the overall performance of the system and is influenced by many variable parameters. Scalability is a major challenge facing the successful deployment of sensor network systems as their size and spatial spread are expected to increase over their lifetimes. The effect of scalable mechanisms on the overall operational performance should also be taken into account. As the network increases in size, scalable mechanisms will have a decreasing effect on the performance of the entire network. Networks with scalable mechanisms are more likely to be operational for longer than those with non-scalable mechanisms. Scalability describes the growth potential and resulting capacity of the network. Scalable networks are those that can expand in terms of node additions and spare resources to support this growth; expanding the network size should not drastically affect performance. The influence of the network size on the average performance is considered. Given a certain network size, there is an optimal resource allocation. Proper resource allocation forms the boundaries of acceptable performance for a certain network size [?].

*3) Energy Efficiency:* Due to the deployment of wireless sensor networks in remote and sometimes harsh environments where human intervention is either expensive or even impossible, energy efficiency has become a very important performance metric in these networks. One of the main performance weaknesses in wireless sensor networks arises from the limited battery capacities of the nodes, also referred to as motes, which are used as sensor nodes. The limited battery capacity of the sensor nodes is mainly due to the fact that they must be extremely small in size, i.e., micro-sized, in order to ensure the miniaturization and low cost of the final wireless sensor network solution. The limited battery capacity implies that the lifetimes of many wireless sensor networks are also limited, especially if the batteries cannot be recharged. Ideally, autonomous sensor networks should sustain themselves for the duration of the mission. This is a fundamental environmental requirement. For sensor networks deployed in situations of catastrophic failure, such as after an earthquake, networks should replenish themselves in order of the expected return cycle time. Using energy scavenging methods, battery pack replacement must be considered the rare exception rather than the rule. Energy scavenging methods extend the operational lifetime of the network provided the first node is powered in a single occupancy dwelling.

Energy efficiency refers to the amount of energy dissipated by a sensor node to perform a task. By improving a sensor network's energy efficiency, the network lifetime increases while maintenance costs decrease, which is an increase in quality of service. The quality of service of the wireless sensor network is influenced by the performance of energyefficient techniques. Achieving a wireless sensor network with very high energy-efficient performance is critical to the development of a suitable self-organization. Moreover, achieving CD, DS, and adaptive modulation requires an energyefficient network. Low energy-consuming radio circuits and adaptive energy-efficient methods to lower processor active and idle current drain were examined in order to increase the network's lifetime. In conclusion, energy-efficient wireless sensor networks offer users improved satisfaction and better performance while reducing network operational expenditures; they are environmentally friendly.

*A. Case Studies and Simulation Results*

This section presents several case studies and some simulation results, the aim of these examples being to assess the effectiveness of our self-organizing features for performance improvement in WSNs. All of these case studies aim to evaluate the practical application of concepts discussed in the previous sections for real-world implementation. Proper simulation methodologies have all been used for evaluating the performance of the network in some context; another aim has been to increase the confidence in these results. In the case studies, we have used either the simulation results or practical data from them to illustrate the impact on the key performance metrics that we have outlined in the previous section.

For example, the effectiveness of one of our self-organizing features can be shown using information that we have received from measuring reliability, scalability, and energy efficiency. In this theme, related solutions we have tried to follow the recommendations. The reply deals with details of the solution and provides a strategy for empirical validation, in that case via detailed simulations.

In conclusion, we believe that different simulation techniques can lead to a better understanding of the solution's effectiveness. In the four examples, some simple case studies (without detailed effect on the schedule) have been given to show the effect of the proposed self-organizing solutions. Furthermore, some simulation results have also been reported to assess the effectiveness of the self-organizing solutions at improving the performance of WSNs. These results are also considered to be of interest to both the academic and industrial communities, providing an opportunity to further evaluate our claims. The examples demonstrate that by varying the way in which the sensors are deployed, through clustering or with probabilities based on the energy levels, it is possible to provide preventive solutions that hold up better against faulty nodes. These results underline the effectiveness of using selforganizing features for WSN.

## IV. CONCLUSION AND FUTURE DIRECTIONS

In this paper, we have presented the performance evaluation of self-organizing features in an advanced wireless sensor networking simulator. More specifically, we have reported on the reliability, scalability, and energy efficiency of four selforganizing mechanisms targeting the solution to two widely studied and difficult problems: that of providing time synchronization and encryption for large wireless sensor networks. The reliability metric provides insight into whether and with what probability the WSN will find a solution within the network. Results demonstrated that the logically centralized, cost-delayed approach resulted in nearly certain solutions. However, the energy required to reach a solution was less in the case of using self-organizing mechanisms. The scalability metric provided insight into how the number of nodes in a WSN affected the success of the problem. Results showed deteriorating performance for available solutions as network size increased. The energy metric provided insight into the shelf life of the WSN. Results advocated that mechanisms must be configured to control their network life cycle to also satisfy connectivity requirements.

In conclusion, we have successfully adopted a method to present performance results with notable deficiencies. Consequently, with the increased usage of WSN applications and new inter-networking capabilities, interoperability and resource management are some of the future requirements when monitoring network designs. Additionally, the deployment of new mobile devices and communication protocols is emerging. Ongoing research topics on the network include the analysis of PANs/RFID, the effects of nano- and/or micro-sensor-centric interaction, QoS-enabled functionality, self-organization, and security, to name a few. It is hoped that this paper will motivate further study into self-organizing mechanisms, particularly the need for innovation given the seismically changing technological landscape.